%% file: osetmenu.tex
\documentclass[12pt,a4paper]{conference}

\usepackage{fancyhdr}
\usepackage{graphicx,amsmath,amssymb,cite}
\usepackage{multind}
\makeindex{author} \makeindex{subject}

\input MENU07macros.tex

\begin{document}

\Chapter{PHOTOPRODUCTION OF $\omega$ AND $\omega$ IN THE NUCLEAR MEDIUM}
           {Paper's Title}{M. Author \it{et al.}}
\vspace{-6 cm}\includegraphics[width=6 cm]{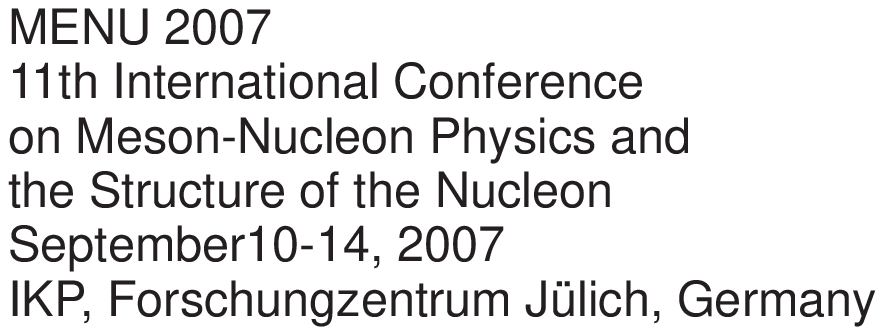}
\vspace{4 cm}

\addcontentsline{toc}{chapter}{{\it N. Author}} \label{authorStart}

\begin{raggedright}

{\it E.Oset$^a$, M. Kaskulov$^b$, H. Nagahiro$^c$, E. Hernandez$^d$ 
and S. Hirenzaki $^e$
\footnote{IFIC, Universidad de Valencia, Spain.}}\index{author}{Author, N.}\\
$^a$Dpto. Fisica Teorica and IFIC\footnote{IFIC, Institutos de Investigacion,
Universidad de Valencia, apdo 22085, 46071 Valencia, Spain.}\\
$^b$ Institut f\"ur Physik, Giessen, Germany\\
$^c$ Research Center for Nuclear Physics, Osaka University, Ibaraki, Osaka\\
$^d$ Facultad de Ciencias, Universidad de Salamanca, Salamanca, Spain\\
$^e$ Department of Physics, Nara Women's University, Nara 630-8506
Japan

\bigskip\bigskip

\end{raggedright}

\begin{center}
\textbf{Abstract}
\end{center}

We reanalyze data from ELSA on $\omega$ production in nuclei, from where claims
of a large shift of the mass were made earlier, which are tied to a certain
election of the background in nuclei, very different in shape to the one on the
proton. The reanalysis shows that the data demand a very large width of the
$\omega$ in the medium, with no need for a shift of the mass, for which the
experiment is quite insensitive. We study possible $\omega$ bound states in the
nucleus and find that, even assuming a small width, they could not be
observed with the present ELSA resolution. Finally we show that, due to the
interplay of background and $\omega$ signal, a two bump structure appears with
the ELSA set up for the $(\gamma,p)$ reaction that should not be misidentified
with a signal of a possible $\omega$ bound state in the nucleus.

\section{Introduction}

The interaction of vector mesons with nuclei has captured for long the 
attention of the hadron community. 
Along these lines, an approach has been followed by the CBELSA/ TAPS
collaboration by looking at the $\gamma \pi ^0$ coming from the $\omega$
decay,  where a recent work \cite{trnka} claims evidence for a decrease of the
$\omega$ mass in the medium of the order of 100 MeV from the study of the
modification of the mass spectra in $\omega$ photoproduction. 
Here we present the reanalysis of the data of \cite{trnka} done in
\cite{muratmass}, where one  
 concludes that the distribution is compatible with
an enlarged $\omega$ width of about 90 MeV at nuclear matter density and no
shift in the mass and at the same time we show the insensitivity of the results
to a mass shift.
We also show results for the $(\gamma ,p)$ reaction searching for possible
$\omega$ bound states in the nucleus concluding that even in the case of a
sufficiently attractive potential and small width no peaks can be seen with the
present experimental resolution of about $50 MeV$  at ELSA. We also discuss the
origin of a two peak structure of the $(\gamma ,p)$ cross section which should
not me misidentified with evidence for an $\omega$ bound state in the nucleus.


\section{Preliminaries}

We consider the  photonuclear reaction 
$A(\gamma,\omega\to \pi^0\gamma)X$ in two
steps - 
production of the $\omega$-mesons and 
propagation of the final states.
In the laboratory, where 
the nucleus with the mass number $A$ is at rest, 
the nuclear total  cross section  of the inclusive reaction $A(\gamma,\omega)X$,
including the effects of Fermi motion and Pauli blocking, plus effects of final
state interaction of the particles produced, can be calculated as
shown in \cite{muratmass}.

The $\omega$-mesons are produced according to their
spectral function $S_{\omega}$ at a local density $\rho(r)$
\begin{eqnarray}
\label{SF}
S_{\omega}({m}_{\omega},\widetilde{m}_{\omega},\rho) = 
\hspace{5cm}\nonumber \\
- \frac{1}{\pi} 
\frac{\mbox{Im}\Pi_{\omega}(\rho)}
{\Big(\widetilde{m}_{\omega}^2-{m}_{\omega}^2
-\mbox{Re}\Pi_{\omega}(\rho)\Big)^2 + 
\Big(\mbox{Im}\Pi_{\omega}(\rho)\Big)^2},
\end{eqnarray}
where $\Pi_{\omega}$ is the in-medium selfenergy of the 
$\omega$. The width of the 
$\omega$ in the nuclear medium is 
related to the selfenergy by 
$\Gamma_{\omega}(\rho,\widetilde{m}_{\omega}) 
= - \mbox{Im}\Pi_{\omega}(\rho,\widetilde{m}_{\omega})/E_{\omega}$.
It includes
the free width $\Gamma_{free} = 8.49$~MeV and an in-medium 
part $\Gamma_{coll}(\rho)$ which accounts for the 
collisional broadening of the $\omega$ due to the quasielastic and
absorption channels. In Eq.~(\ref{SF})
$\mbox{Re}\Pi_{\omega}=2 E_{\omega} \mbox{Re}V_{opt}(\rho)$, where
$V_{opt}(\rho)$ is the $\omega$ nucleus optical potential accounts 
for a possible
shift of the $\omega$ mass in the medium and we shall make some considerations
about it latter on.

We also consider the situation when 
the energy of the incident photon beam  is not fixed but 
constrained in some energy interval
$E_{\gamma}^{\min} < E_{\gamma} < E_{\gamma}^{\max}$, and also take into account
the photon flux produced at the ELSA facility.

\section{The Monte Carlo simulation procedure}

The computer MC simulation proceeds in close analogy to the actual
experiment.
At first,
the multiple 
integral involved in the evaluation of the cross section is carried out using 
the MC integration 
method. This procedure provides
 a random 
point $\vec{r}$ inside the nucleus where the photon collides 
with the nucleon, 
also randomly generated
from the Fermi sea with $|\vec{p}_N| \le k_F(\vec{r})$.
For the sample event in the MC integral
the mass $\widetilde{m}_{\omega}$ of the $\omega$
respects 
the spectral function $S_{\omega}$ at local density $\rho(r)$, 
see Eq.~(\ref{SF}).
Inside the nucleus the $\omega$-mesons moving with the three
momentum $\vec{p}_{\omega}^{\,lab}$ necessarily interact with
the nucleons in their way out of the nucleus. In the MC simulation
the $\omega$-mesons are allowed to propagate  
a distance 
$\delta \vec{L} = \frac{\vec{p}_{\omega}^{\,lab}}{|\vec{p}_{\omega}^{\,lab}|}
\delta L$ and at each step, $\delta L \simeq 0.1$~fm, the 
reaction probabilities for different 
channels like
the decay of the $\omega$
into $\pi^0 \gamma$ and $\pi\pi\pi$ final states, 
quasielastic scattering and in-medium absorption
are properly calculated. Details of
the simulation can be seen in \cite{muratmass}.

 We use the following parameterization for the width,
$\Gamma_{abs} = \Gamma_0 \frac{\rho(r)}{\rho_0}$,
where $\rho_0=0.16$~fm$^{-3}$ is the normal nuclear matter density.

The propagation of pions in nuclei is done using a MonteCarlo simulation
procedure . In their way out
of the nucleus pions can experience the quasielastic scattering or 
can be absorbed. 
The intrinsic probabilities for these reactions
as a function of the nuclear matter density
are calculated using the phenomenological model of 
Refs~\cite{simulation},
which also includes 
higher order quasielastic cuts and 
the two-body and three-body
absorption mechanisms. Details for the present case are described in
\cite{muratmass}.

\section{In-medium $\omega$-meson width and nuclear transparency}

In this section we discuss an extraction of the 
in-medium inelastic width of the
$\omega$ in the photonuclear experiments.
As a measure for the $\omega$-meson width in nuclei we employ the so-called
nuclear transparency ratio
\begin{equation}
\tilde{T}_{A} = \frac{\sigma_{\gamma A \to \omega X}}{A \sigma_{\gamma N \to \omega X}}
\end{equation}
i.e. the ratio of the nuclear $\omega$-photoproduction cross section
divided by $A$ times the same quantity on a free nucleon. 
$\tilde{T}_A$ describes
the loss of flux of $\omega$-mesons in the nuclei and is related to the
absorptive part of the $\omega$-nucleus optical potential and thus to the
$\omega$ width in the nuclear medium.

We have done the MC calculations for the
sample nuclear targets:   ${}^{12}_6$C, ${}^{16}_{8}$O,
${}^{24}_{12}$Mg,  ${}^{27}_{13}$Al, ${}^{28}_{14}$Si,
${}^{31}_{15}$P,   ${}^{32}_{16}$S,  ${}^{40}_{20}$Ca,
${}^{56}_{26}$Fe, ${}^{64}_{29}$Cu,  ${}^{89}_{39}$Y, 
${}^{110}_{48}$Cd,  ${}^{152}_{62}$Sm,  ${}^{208}_{82}$Pb,
${}^{238}_{92}$U. 
 In the following we evaluate the ratio between the nuclear 
cross sections in
heavy nuclei and a light one, for instance  $^{12}$C, since in
this way, many other nuclear effects not related to the
absorption of the $\omega$ cancel in the
ratio, $T_A$.

\begin{figure*}[t]
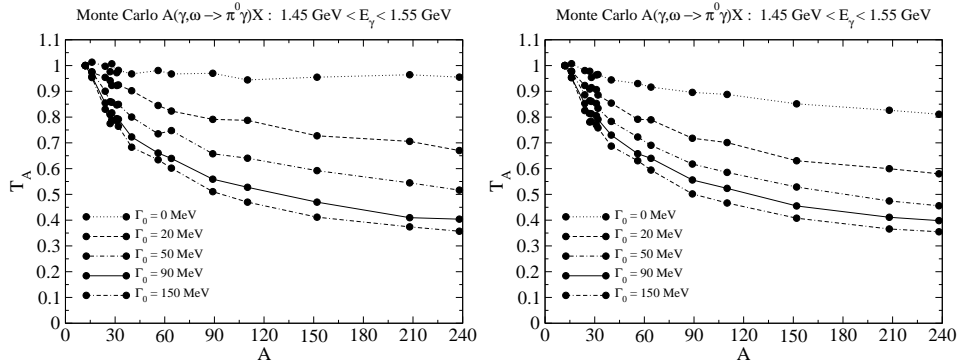

\begin{center}
\includegraphics[clip=true,width=0.45\columnwidth,angle=0.]
{TransparencyMC.eps}
\includegraphics[clip=true,width=0.45\columnwidth,angle=0.]
{TransparencyMCpionCut.eps}
\caption{\label{FigTrMC}  \footnotesize
The result of the Monte Carlo method for the $A$-dependence
of the nuclear transparency ratio $T_A$ without (left panel) and with
(right panel) FSI of outgoing pions. A lower cut $T_{\pi} > 150$~MeV 
on the kinetic energy of the outgoing pions has been used to suppress 
the contribution of the distorted events due to FSI.
The incident photon beam was constrained in the range
$1.45~\mbox{GeV}< E_{\gamma} < 1.55~\mbox{GeV}$.
 The carbon $^{12}$C
was used as the reference target in the ratio of the nuclear cross sections.
With
$\Gamma_{abs} = \Gamma_0 \frac{\rho(r)}{\rho_0}$, where $\rho_0$ is 
the normal nuclear matter density, the dotted, dashed, dash-dotted,
solid and dash-dash-dotted  
curves 
correspond to $\Gamma_0 = 0$~MeV, $\Gamma_0=20$~MeV, $\Gamma_0=50$~MeV, 
$\Gamma_0=90$~MeV and $\Gamma_0=150$~MeV, respectively. 
}
\end{center}
\end{figure*}



The results of the MC calculation
for the $A$-dependence
of the nuclear transparency ratio $T_A$ are presented in Fig.~\ref{FigTrMC}.
The incident photon beam was constrained in the range
$1.45~\mbox{GeV}< E_{\gamma} < 1.55~\mbox{GeV}$ - a region which is considered
in the analysis of the CBELSA/TAPS
experiment~\cite{trnkathesis,Kotulla:2006wz}. 
In Fig.~\ref{FigTrMC} (left panel) we show the results for the
transparency ratio when the collisional broadening and FSI of the $\omega$
are taken into account but without FSI
of the pions from $\omega \to \pi^0 \gamma$ decays inside the nucleus. The right
panel corresponds to considering in addition the FSI of the pions.

By using these results and taking into account
the preliminary results of CBELSA/TAPS experiment~\cite{Kotulla:2006wz} 
we get an estimate 
for the $\omega$ width $\Gamma_{abs} \simeq 90 \times
\frac{\rho(r)}{\rho_0}~\mbox{MeV}$.
This estimate must be understood as an average over the 
$\omega$ three momentum.

\section{In-medium $\omega$-meson mass and CBELSA/TAPS experiment}

The first thing one should note is that the $\omega$ line shape reconstructed
from $\pi^0\gamma$ events strongly depends
on the background shape subtracted from the bare $\pi^0\gamma$ signal.
In Ref.~\cite{trnka} the shape of the background was chosen such that it
accounted for all the experimental strength at large invariant masses.
This choice was done both for the elementary $\gamma p \to  \pi^0 \gamma p$
reaction as well as for nuclei. As we shall show, this choice of background in
nuclei implies a change of the shape from the elementary reaction to that in
the nucleus for which no justification was given. We shall also show
that when the
same shape for the background as for the elementary reaction is chosen, 
the experiment in nuclei shows strength at invariant masses higher than 
$m_{\omega}$ where the choice of~\cite{trnka} necessarily produced no
strength. We will also see that the experimental data can be naturally
interpreted in terms of the large in-medium $\omega$ width discussed above
without
the need to invoke a shift in the $\omega$ mass in the medium. 

\begin{figure}[t]
\begin{center}
\includegraphics[clip=true,width=0.90\columnwidth,angle=0.]
{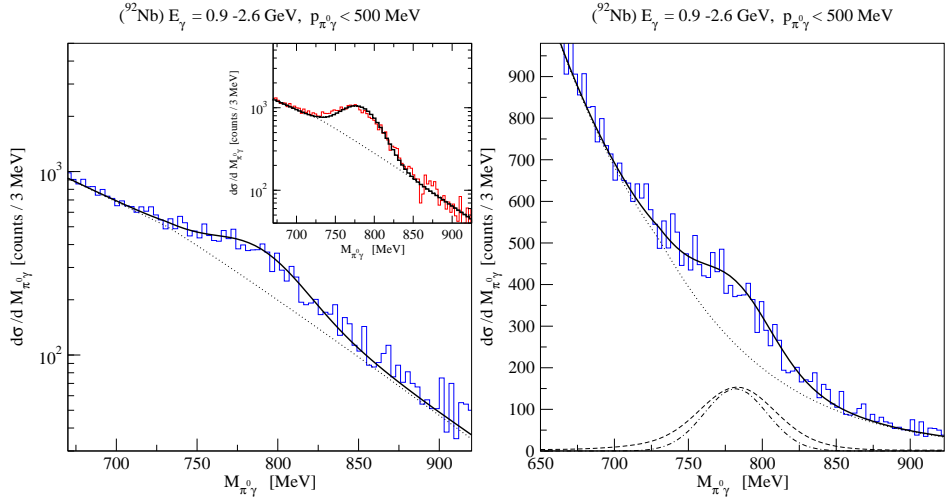}
\caption{\label{PRL}  \footnotesize
Left panel: Invariant mass spectra reconstructed from the $\pi^0\gamma$
events in the $(\gamma,\pi^0\gamma)$ reaction from Nb target (solid curve). 
The experimental data are from Ref.~\cite{trnka}. 
The incident photon beam has been constrained
in the range $0.9$~GeV $< E_{\gamma}^{in} <$ $2.6$~GeV.
Dotted curve is an uncorrelated $\pi^0\gamma$
background (see the text). 
Right panel: Same but in linear scale. 
The dashed and dash-dash-dotted curves correspond to the
$\omega \to \pi^0\gamma$ events with and without the kinematic cut
$|\vec{p}_{\pi^0\gamma}| < 500$ MeV,
respectively. The normalization without cut is arbitrary.
The solid line corresponds to the sum of the
background and the dashed line.
Inset (left panel): The $\pi^0\gamma$ invariant mass spectra in the elementary
$p(\gamma,\pi^0\gamma)p$ reaction. 
Same background line shape (dotted curve) as for the Nb target has been used.
The solid line is the sum of the
background and $\omega \to \pi^0\gamma$ events.}
\end{center}
\end{figure}

In Fig.~\ref{PRL}  
we show the experimental data (solid histogram) for the 
$\pi^0\gamma$ invariant mass spectra in the reaction
$(\gamma,\pi^0\gamma)$~\cite{trnka} from $^{92}_{41}$Nb target. 
The inset (left panel) corresponds to the $\pi^0\gamma$ spectra from the
hydrogen target. 
In our MC calculations the incident photon beam has been constrained
in the range $0.9$~GeV $< E_{\gamma}^{in} <$ $2.6$~GeV. 
The higher momentum cut 
$|\vec{p}_{\pi^0\gamma}| = |\vec{p}_{\pi^0}+\vec{p}_{\gamma}|< 500$~MeV
on a three momentum of the $\pi^0 \gamma$ pair
was imposed as in the actual experiment. First, we use the hydrogen target,
see inset in Fig.~\ref{PRL} (left panel), 
to fix the contribution of the uncorrelated $\pi^0\gamma$
background  (dotted curve) which together with
the $\pi^0\gamma$ signal from $\omega \to \pi^0\gamma$ decay,
folded with the Gaussian experimental resolution of 55 MeV as 
in Ref.~\cite{trnka}, gives a fair
reproduction of the experimental spectra. Then we assume the same 
shape of the $\pi^0\gamma$ 
background in the photonuclear reaction.
The weak effect of the FSI of the pions found in the calculation, with the cuts
imposed in the experiment, strongly supports this assumption.

In the following we use the $\omega$ inelastic 
width of $\Gamma_0 = 90$~MeV
at $\rho_0$.
The exclusive
$\omega \to \pi^0\gamma$ MC spectra is shown by the dashed curve 
(right panel). 
The solid curve is the reconstructed
$\pi^0\gamma$ signal after applying the cut on $\pi^0\gamma$ momenta and
adding
the background fixed when using the hydrogen target (dotted curve). 
Note that the shape of the exclusive $\pi^0 \gamma$ signal without applying 
a cut on $\pi^0\gamma$ momenta
(dash-dotted curve) is dominated by the experimental resolution 
and no broadening of the $\omega$ is observed. This is in agreement
with data of Ref.~\cite{trnka}.
But applying the cut one increases
the fraction of in-medium decays coming from the interior of the nucleus 
where the spectral function is rather broad
and as a result the broadening of the $\pi^0\gamma$ signal with respect
to the signal (without cut) can be well seen.
The resulting MC spectra (solid curve) 
shows the accumulation of the $\pi^0\gamma$ events from the left and right 
sides
of the mass spectra, and it is consistent both with 
our choice of the uncorrelated $\pi^0\gamma$
background and experimental data.

We have also done the exercise of seeing the sensitivity of the results to 
changes in the mass. As shown in \cite{muratmass}, a band corresponding to
having the $\omega$ mass in between $m_{\omega}\pm 40 \rho/\rho_0$~MeV is far
narrower than the statistical fluctuations.
In other words, this 
experiment is too insensitive to changes in the mass to be used for a precise
determination of the shift of the $\omega$-mass in the nuclear medium.
We should also note that the peak position barely moves since it is
dominated
by the decay of the $\omega$ outside the nucleus.

\section{Production of bound $\omega$ states in the ($\gamma$,p)  
reaction}

Here we evaluate the formation rate of  $\omega$ bound states in the 
nucleus by means of the ($\gamma$,p) reaction.  We use the  
Green function method \cite{NPA435etc} to calculate the cross  
sections for $\omega$-mesic states formation as described in
Refs.~\cite{NPA761} in   
detail.
The theoretical model used here is exactly same as that used in these
references.

The $\omega$-nucleus optical potential is written here as 
$V(r) = (V_0 + iW_0  ) \frac{\rho(r)}{\rho_0}$, 
where  $\rho(r)$ is the nuclear experimental density for which we take 
 the two parameter Fermi  distribution.
We consider three cases of the potential strength as: $(V_0,W_0) = -(0,50)$,
$-(100,50)MeV$ and $-(156,29) MeV$.
The last of the potentials is
obtained by the linear density approximation with the scattering length
$a=1.6 + 0.3 i$ fm \cite{klingl2}.
This potential is strongly attractive with weak
absorption and hence should be the ideal case for the formation
of $\omega$ mesic nuclei.
No $\omega$ bound states are expected for the first  potential 
which has only an absorptive part.
The second potential  has a strong attraction 
with the large absorptive part as indicated in
Ref.~\cite{trnkathesis}.
For the first two potentials we find no visible peaks in the
spectrum since the width is so large.
For the third potential we observe peaks but they are washed out when folded 
with the experimental resolution of about $50 MeV$ of ELSA.

\section{Monte Carlo simulation of the reaction of the $(\gamma,p)$ reaction}

We next apply the MonteCarlo simulation explained above to describe the 
$(\gamma,p)$ reaction studied at ELSA.
Because our MC calculations represent complete event simulations
it is possible to take into account the actual experimental 
acceptance of ELSA~\cite{trnkathesis} (see details in \cite{hidekoatoms}).

We start our MC analysis with the cross section of 
the elementary reaction $\gamma p \to \omega p \to \pi^0\gamma p$. With this we
determine the cross section for $\omega$ formation and follow the fate of the
protons at the same time.

\begin{figure*}[t]
\begin{center}
\includegraphics[clip=true,width=0.70\columnwidth,angle=0.]
{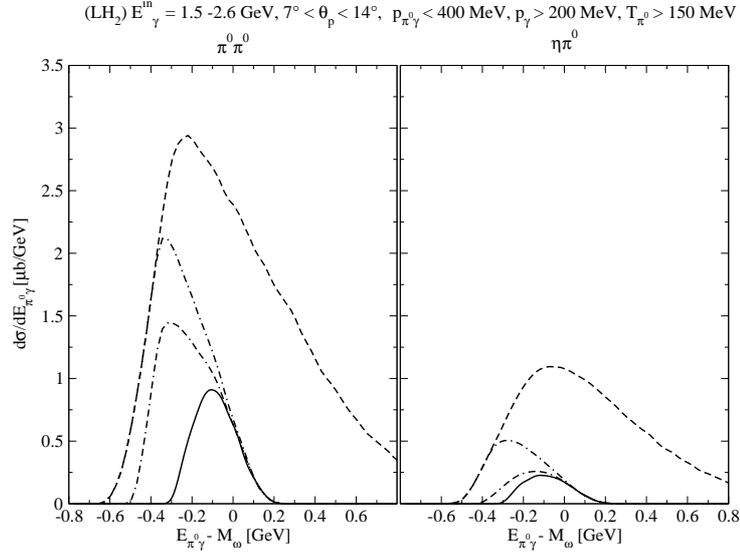}
\caption{\label{FigNewBGR} \small
The differential cross section $d\sigma/dE_{\pi^0 \gamma}$ 
of the reactions $\gamma p \to \pi^0 \pi^0  p$ (left panel)
and $\gamma p \to \pi^0 \eta  p$ (right panel) followed by the decay
$\pi^0(\eta) \to \gamma \gamma$
as a function of the
$E_{\pi^0\gamma}-m_{\omega}$ where $E_{\pi^0\gamma}=E_{\pi^0}+E_{\gamma}$.
 The following cuts were imposed: 
$E_{\gamma}^{in}= 1.5 \div 2.6$~GeV and $7^{\circ} < \theta_p < 14^{\circ}$
(dashed curves);
$E_{\gamma}^{in}= 1.5 \div 2.6$~GeV, $7^{\circ} < \theta_p < 14^{\circ}$
and $|\vec{p}_{\pi^0}+\vec{p}_{\gamma}| < 400$~MeV (dash-dotted curves);
plus  the cut $T_{\pi^0} > 150$~MeV (dash-dash-dotted curves)
 and plus the cut $|\vec{p}_{\gamma}| > 200$~MeV (solid curves).
}
\end{center}
\end{figure*}

There are  also sources of background like from $\gamma p \to \pi^0
\pi^0 p$, or $\gamma p \to \pi^0 \eta p$, where one of the two photons from
the decay of the $\pi^0$ or the $\eta$ is not measured. 
 We show in Fig.~\ref{FigNewBGR} 
the cross section 
$d \sigma /d E_{\pi^0 \gamma}$ coming from the 
$\gamma p \to \pi^0\pi^0 p$ reaction 
followed by the decay $\pi^0 \to \gamma \gamma$ of either of the 
$\pi^0$ (left panel)
and from the $\gamma p \to \pi^0\eta p$ reaction followed by the decay 
$\eta \to \gamma \gamma$ (right panel). 
  As one can see, the
contribution from the $\pi^0 \pi^0$ 
photoproduction to the background is the dominant
one among the two.  The important thing, thus, is that these two 
sources of background, with the cuts imposed,  produce a background 
peaked at -100 MeV. For the exclusive $\pi^0 \gamma$ 
events coming from $\gamma p \to \omega p \to \pi^0\gamma p$ 
an experimental resolution of $50$~MeV was imposed, 
see Ref.~\cite{trnka}. 
We obtain a factor of two bigger strength
at the $\omega$ peak than at the peak from the $\gamma p \to \pi^0 \pi^0 p$
background. Experimentally, this seems to be also
the case from the preliminary data of CBELSA/TAPS,

\begin{figure*}[t]
\begin{center}
\includegraphics[clip=true,width=0.70\columnwidth,angle=0.]
{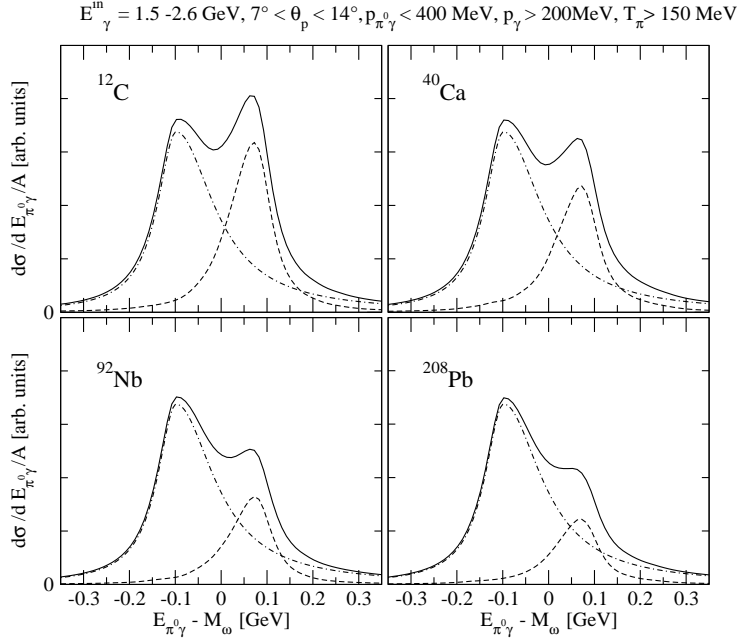}
\caption{\label{FigEMnuclear} \small
The differential cross section $d\sigma/dE_{\pi^0\gamma}$ 
of the reaction $A(\gamma,\pi^0\gamma)X$
as a function
of $E_{\pi^0\gamma} - m_{\omega}$  from $^{12}$C, $^{40}$Ca, $^{92}$Nb and
$^{208}$Pb nuclear targets. The reconstructed exclusive events from the 
$\omega \to \pi^0 \gamma$ decay are
shown by the dashed curves.  The $\pi^0 \gamma$ background is shown by
the dash-dotted curves. 
The sum of the two contributions is given by the solid curves.
The following cuts were imposed: 
$E_{\gamma}^{in}= 1.5 \div 2.6$~GeV, $7^{\circ} < \theta_p < 14^{\circ}$,
$|\vec{p}_{\pi^0}+\vec{p}_{\gamma}| < 400$~MeV, $|\vec{p}_{\gamma}| > 200$~MeV 
and $T_{\pi} > 150$~MeV. The exclusive $\omega \to \pi^0 \gamma$ signal 
has been folded with the 50 MeV experimental resolution.
All spectra are normalized to the corresponding 
nuclear mass numbers $A$.} 
\end{center}
\end{figure*}


In the following we assume that the inclusive
$\pi^0 \gamma$ background scales with respect to 
the target nucleus 
mass number $A$ like $\sigma_A \simeq A \, \sigma_{elem}$.
But this is not
the case for the exclusive $\pi^0 \gamma$ events coming from the decay of the
$\omega \to \pi^0 \gamma$, since 
the rather strong absorption
of the $\omega$ inside the nucleus changes the scaling relation
and $\sigma_A(\omega \to \pi^0 \gamma) \simeq A^{\alpha} \, 
\sigma_{elem}(\omega \to \pi^0 \gamma)$,
 where the attenuation parameter $\alpha < 1$.

In Fig.~\ref{FigEMnuclear} 
we show the result of the MC simulation for the 
$E_{\pi^0 \gamma} - m_{\omega}$ 
spectra reconstructed from the $\pi^0$ and $\gamma$ events. The calculations
are performed 
for the sample nuclear targets $^{12}\mbox{C}$, $^{40}\mbox{Ca}$, 
$^{92}\mbox{Nb}$ and $^{208}\mbox{Pb}$. The kinematic and acceptance
cuts discussed before  have been already imposed. 
The MC distributions are
normalized to the nuclear mass number $A$. The solid curves correspond
to the sum of the inclusive $\pi^0 \gamma$ background (dash-dotted curve), 
 and the exclusive $\pi^0 \gamma$ 
events coming from the direct
decay of the $\omega\to \pi^0 \gamma$. The contributions of the exclusive
$\omega \to \pi^0 \gamma$
events are shown by the dashed curves. We note a
very strong attenuation of the
$\omega\to \pi^0 \gamma$ signal with respect to the background contribution
 with increasing  nuclear mass number $A$.
 This is
primary due to the stronger absorption of the $\omega$-mesons 
with increasing nuclear matter density.
  The former exercise indicates that given the particular combination of 
$\pi^0 \gamma$ from an uncorrelated background and from  $\omega$ decay, and
the different behaviour of these two sources in the $\pi^0 \gamma$ production in
nuclei, a double hump structure is unavoidable in nuclei with this set up, and 
one should avoid any temptation
to associate the lower energy peak to a possible bound state in the nucleus. 

\section{\label{Summary} Conclusions}
 The studies done in \cite{muratmass} and \cite{hidekoatoms} show that: 1) The
 ELSA results on inclusive $\omega$ production in nuclei can be interpreted in
 terms of a large $\omega$ width in the medium without the need of a mass shift.
  2) The results are very insensitive to a mass shift in  matter. 3) With the
  large medium $\omega$ width derived from the ELSA data no visible peaks for
  $\omega$ bound states are seen, even with hypothetical large $\omega $
  binding. 4) Even in the hypothetical case of small widths, the possible
  $\omega$ bound states would not be resolved with the present ELSA resolution.
  5) When looking at the $(\gamma,p)$ reaction with the present ELSA
  experimental set up, a double hump structure appears in the calculation from
  the interplay of the $\omega$ signal and the background. The peak at lower
  energies is related to the background, with the cuts imposed, and should not
  me misidentified with a possible $\omega$ bound state in the nucleus.

\section*{Acknowledgments}  
This work is partly supported by DGICYT contract
number BFM2003-00856, the Generalitat Valenciana, the projects FPA2004-05616
(DGICYT) and SA016A07 (Junta de Castilla y Leon).
 This research is  part of the EU Integrated
Infrastructure Initiative  Hadron Physics Project under  contract number
RII3-CT-2004-506078.

\section{References}


\end{document}

%% file: MENU07macros.tex
\newcommand{\beq}{\begin{equation}}
\newcommand{\eeq}[1]{\label{#1}\end{equation}}
\newcommand{\eeqn}{\end{equation}}

\newcommand{\beqa}{\begin{eqnarray}}
\newcommand{\eeqa}[1]{\label{#1}\end{eqnarray}}
\newcommand{\eeqan}{\end{eqnarray}}

\let\bar=\overbar

\newcommand{\Dslash}{\not{\hbox{\kern-4pt $D$}}}
\newcommand{\dslash}{\not{\hbox{\kern-2pt $\del$}}}

\newcommand{\msb}{{\bar{\ssstyle M \kern -1pt S}}}

